# Transparency and Coordination in Peer Production

LAURA DABBISH, JASON TSAY, AND JIM HERBSLEB, Carnegie Mellon University
COLLEEN STUART, Johns Hopkins University

## 1. INTRODUCTION

Transparency, or the accurate observability of an organization's low-level activities, routines, and outputs (Bernstein, p. 181), provides powerful capabilities for coordination. Transparency in a collocated work environment (e.g. Teasley et al., 2000) can reduce coordination conflicts (Carlile, 2002; Bechky, 2006), and enable the transfer of best practices or new ideas (Tortoriello & Krackhardt, 2010). Under the right conditions, transparency can positively impact on learning and performance (Argote & Ingram, 2000).

Coordination is particularly challenging when work is organized as commons-based peer production (CBPP) (Benkler, 2002). Having no recourse the mechanisms of hierarchy or markets, the more well-understood means of allocating effort to tasks, CBBP relies on individuals to identify opportunities to work and to coordinate their work with others (Benkler, 2002). Increasingly, many such communities are relying on digitally transparent work environments to enable self-coordination, even at very large scales (Herbsleb & Mockus, 2003; Mockus, Fielding, & Herbsleb, 2002). Understanding digital transparency, the organizational processes it supports, and the requirements and limits of its effectiveness has important implications for grasping its potential in a wider range of contexts.

In contrast to collocated transparent work environments, where information is often transferred through observation and overhearing, new electronic work environments have the potential to support a different form of digital transparency because of the richly detailed data they store about artifacts, activities on those artifacts, and people (Stuart, et al., 2012). By virtue of this detailed information, transparency potentially extends far beyond the immediate physical workspace, both in space and in time. Activities anywhere in the world are potentially just as visible as those undertaken locally. Activities from the recent or distant past are as easy to "see" as those in the current moment. Transparency can have much greater scope and permanence. Moreover, in these environments, both the content of work and the process used to achieve it are made visible, and even the subtlest of actions taken in the digital space can be accessible to an audience that could include everyone with access to the environment. Research on technologies that provide cross-workspace awareness of worker activities suggests that this information can reduce integration time (Sarma, Noroozi, & van der Hoek, 2003) and support broad awareness of important changes or events (Kraut & Streeter, 1995).

Greatly expanding the scope and detail of work transparency presents potential dangers as well as potential benefits. Plentiful information creates a scarcity of attention (Simon, 1971), and information overload can become a serious issue. Moreover, observers may have difficulty making sense of actions when the behavior is not directed at them, they are unclear of the actor's intent, or they lack the context-specific knowledge to accurately make sense of the event (Seeley-Brown & Duguid, 2000). Finally, an ever-present audience may be detrimental to work when hiding may be necessary for experimentation or to streamline workflows (Bernstein, 2012).

Despite the rapid incorporation of digital transparency into peer production work environments, we know little about how work is actually coordinated in these environments. When information is plentiful, and potentially overwhelming, how is it extracted and applied to coordinate work? People are social creatures and make inferences about others from what they observe (e.g., Goldstein & Cialdini, 2008), but it is unclear what people are able to infer from such a collection of information, and how these inferences help them carry out their collective work.

The literature on coordination hints at a couple of ways that transparency could influence coordination in organizations. Coordination can be defined as "managing dependencies between activities" (Malone & Crowston, p. 90), Whenever an individual's task is interdependent with activities of another individual, that activity may be influenced by the actions of other individuals. There is an opportunity for those actions to be organized, or coordinated, to manage their potential influence on each other. By increasing the information that organizational members are able to access, transparency influences an individual's tacit knowledge, and the shared knowledge that held by members of an organization. Since the history of work artifacts and actions on them are available to everyone, the environment provides common ground (Brennan and Clark, 1990) for any discussion of the work.

A key attribute of a transparent work environment is that individuals have unrestricted access to vast amounts of information concerning people and their activities, irrespective of whether they are physically collocated. In a peer production setting, how do participants coordinate their work when individuals have unparalleled access to information embedded in work artifacts? How do individuals extract and make use of this knowledge? Further, how does the awareness that their own activities are visible, and documented for perpetuity, influence their own behavior? To address these questions, we used qualitative research methodologies to better understand the processes underlying coordination in a transparent work environment.

In order to understand how work is coordinated in transparent work environments, we conducted a qualitative field investigation of a community of software developers who use a website called GitHub (GitHub.com) to host their open source software (OSS) repositories. GitHub is an iconic example of a transparent work environment. The site integrates a number of social features that make unique information about users and their activities visible within and across open source software projects. We found that transparency supported coordination within the community by alerting members to changes and potential conflicts earlier in a projects development trajectory. The design of the environment gave them timely opportunities to adjust their own work or intervene and attempt to address the conflict. Two important characteristics enabled coordination: a developer's experience within the environment allowed him to derive context-specific knowledge from the trace information, and awareness of being observed by an audience of peers led developers to write code in a way that made it was legible to other users. In the next section we briefly describe our study and results.

## 2. STUDY: COORDINATION IN A TRANSPARENT ENVIRONMENT

We conducted a qualitative study of software developers in the context of a system called GitHub that integrates social networking functionality closely with the development environment itself. GitHub is a web-based hosting service for OSS projects that use the Git version control system with a number of web services that provide social media functionality. Compared to other software hosting sites because in addition to allow users to control revisions to their projects, it has social networking functionality that provides users with real-time updates on what others are working on. As of September 2011 GitHub hosted over 2 million code repositories and had one million registered users (GitHub.com).
We conducted a series of semi-structured interviews with developers on GitHub. To obtain a diversity of information about use, we sampled both peripheral users with projects that had only a few watchers and heavier users with more popular projects. We also sampled across users working on the site in a voluntary way versus as part of the work. In our interviews we asked developers to walk us through their last session on the site, thinking aloud about the displayed information on the site, their interpretation of it, and what they did. We then transcribed our interviews and isolated incidents of inferences made on displayed site information and subsequent actions. We open coded these incidents for a subset of five interviews, and then applied our codes to the rest of the interviews to refine the coding scheme. We analyzed in detail the information that's available on the site, and the social inferences that people made based on that information

### 1.1   Findings: Coordination and Pre-Coordination

Our research goal was to examine how coordination takes place in transparent work environments. Our analysis showed that in a transparent work environment people were able to extract knowledge about content people and interactions, from visible work artifacts, that affected the way they coordinated their work. Coordination was enabled by a series of pre-coordination behaviors. These included: 1) continuously monitoring the evolution and trajectory of current and future dependent projects, as well as 2) making the intention behind work activity legible through construction of actions and 3) notifying dependent projects of important decision points and opportunities for mutual adjustment. We label these behaviors "pre-coordination" because they supported early identification of potential conflicts or lowered the likelihood of conflicts occurring. Through our analysis, we identified several enabling conditions for pre-coordination behaviors. These included the presence of an observing audience, and how it affects the quality of the information that is encoded within the site. The process of scouting potential dependencies or monitoring existing ones, involved reading out knowledge from visible information in the environment. We found that work activity and process knowledge was captured in visible traces in the environment, because of the permanence and granularity of activity traces.

Pre-coordination scouting and monitoring behaviors involved the extraction of task and activity knowledge from work artifacts. This process was notable because in many cases this knowledge transfer was able to occur without direct interaction. Developers were able to acquire knowledge through their inferences about traces of visible activity available in the environment (code, comments around code, work history). Process knowledge about how work is conducted is typically difficult to acquire without interaction (Nonaka, 1994). This type of knowledge can't be easily documented. We observed, however, that in a transparent environment users could often infer information thought of as typically tacit information from work artifacts and visible activity history.

The knowledge users were able to extract from the environment allowed them to coordinate with producer and consumer projects in unique ways. They could anticipate changes in interdependent projects well in advance, and interact directly with project owners in response to problematic changes observed. The transparent work environment supported access to tacit knowledge that supported a form of social sensemaking - managing dependencies – using the history of the repository as a resource, as a way of knowing how to approach the owner of a repository, or who to contact.

Like other coordination mechanisms, transparency promotes accountability and predictability and common understanding.  An audience for work activity led to accountability for changes to interdependent projects, and construction of work traces that were understandable. Predictability and understanding stemmed from the deep level of informational access that people got from the environment. However, unlike other coordination mechanisms, where the informational benefits of transparency stem from familiarity through interaction, this knowledge is transferred asynchronously. Our results have implications for the design of transparent work environments for commons-based peer production.